\documentclass{moriond}

\def\be{\begin{equation}}
\def\ee{\end{equation}}
\def\bea{\begin{eqnarray}}
\def\eea{\end{eqnarray}}

\newcommand{\Photo}{}

\begin{document}
\vspace*{4cm}
\title{LOW MASS WIMP SEARCH WITH EDELWEISS-III: FIRST RESULTS}

\author{ Thibault DE BOISSI\`{E}RE\footnote{thibault.main-de-boissiere@cea.fr} (for the EDELWEISS collaboration) }

\address{CEA, Centre d'\'{E}tudes Saclay, IRFU, 91191 Gif-Sur-Yvette Cedex, France}

\maketitle\abstracts{
We present the first search for low mass WIMPs using the Germanium bolometers of the EDELWEISS-III experiment. Upgrades to the detectors and the electronics enhance the background discrimination and the low energy sensitivity with respect to EDELWEISS-II. A multivariate analysis is implemented to fully exploit the detector's potential, reaching a sensitivity of 1.6 $\times 10^{-5}$ pb for a WIMP mass of 7~GeV/c$^2$ with a fraction of the data set, unblinded for background modeling and analysis tuning.}

\section{The EDELWEISS experiment}

A variety of observations (see \cite{bib:Bertone} for a review) point to the existence of cold, non baryonic dark matter which would make up to 27\% of the content of the universe~\cite{bib:Planck}. After decades of experimentations, its nature remains poorly constrained. One well-motivated candidate (referred to as a Weakly Interacting Massive Particle or WIMP) arises from Beyond Standard Model theories like SUSY. Its mass and weak cross section naturally provide the observed relic density. If WIMPs exist, they should be found in the galactic halo and are expected to scatter off nuclei on Earth. This motivates direct detection experiments which search for nuclear recoils in massive detectors. The experimental challenge (background rejection at $\sim$ keV energies) requires the use of extremely pure materials in clean environments. 

\begin{figure}[!ht]
\label{spectrum_data}
\centering
\includegraphics[height=6cm,width=0.49\textwidth]{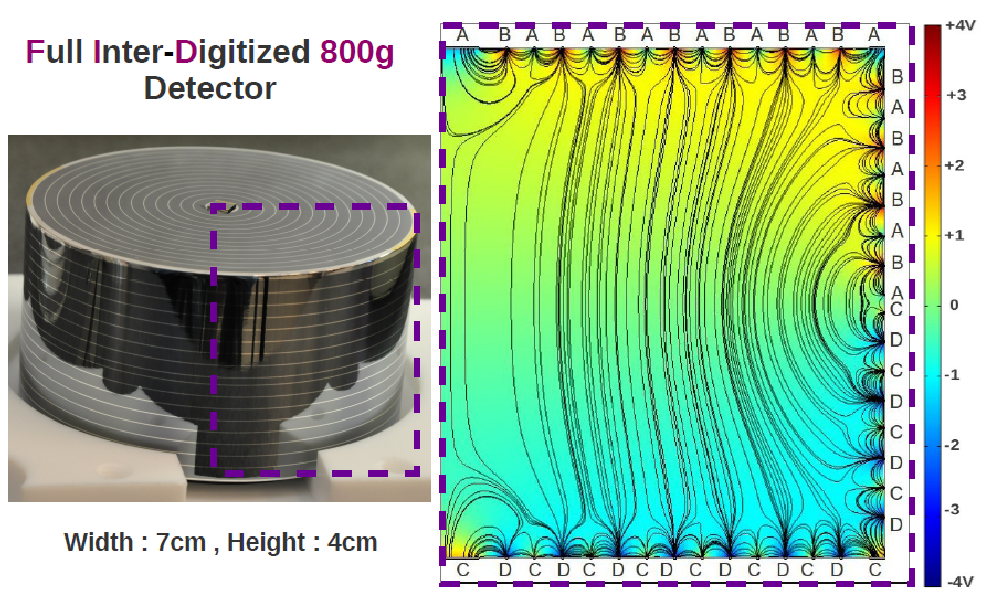}
\includegraphics[height=6cm,width=0.49\textwidth]{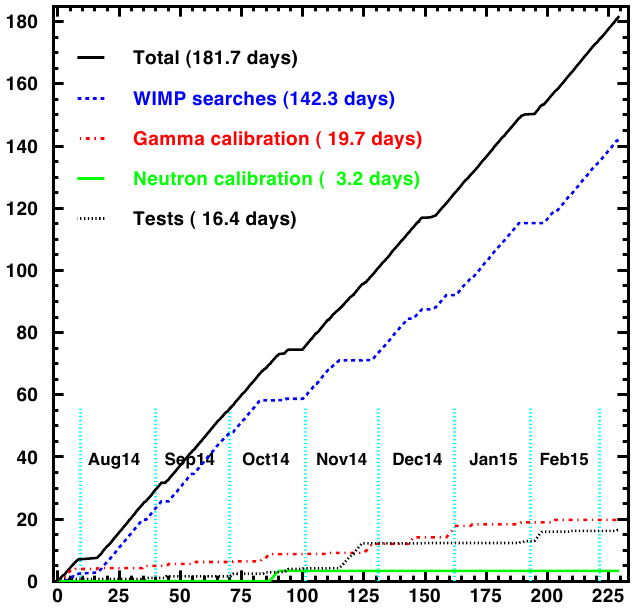}
\label{fig:FID}
\caption{{\bf Left:} EDELWEISS-III Fully Inter-Digitized (FID) detector. The black lines show the electric field within the detector. {\bf Right:} EDELWEISS-III data acquisition progress since summer 2014.}
\end{figure}

~\\The EDELWEISS experiment operates germanium bolometers at very low temperatures (18~mK) in the Underground Laboratory of Modane. The detectors are protected from external radioactivity by lead and polyethylene (PE) shields. An active veto monitors cosmic muons in order to reject muon-induced neutrons, which can mimic the WIMP signal. Each detector is equipped with a set of interleaved electrodes and thermometers. These sensors measure the ionization and phonon signals triggered by incoming particles. The comparison of the two signals allows a separation of nuclear recoils from electron recoils induced for instance by $\beta$ and $\gamma$ radioactivity. The interleaved design modifies the field lines near the surface (see Fig.~\ref{fig:FID}). This allows us to define a fiducial volume for each detector and to reject near-surface interactions using the information on the veto electrode. 
\newline The current setup includes some notable improvements over the previous phase of the experiment. A new PE shielding has been inserted between the bolometers and the electronics, while the copper used in the thermal shields has been replaced by much purer NOSV copper.  The cryogenics have been upgraded as well: thermal machines are now placed outside the shields, allowing microphonics reduction. By replacing the feedback resistances with mechanical relays, the ionisation read-out was improved, yielding a 30\% better baseline resolution. The detector design has evolved as well: detectors now carry interleaved electrodes on the lateral surfaces. This increases the share of the fiducial volume to 75\%, up from 40\% and allows better rejection. The detectors are also more massive: 800~g up from 400~g.

\section{Low mass WIMPs}

Recent excesses of events have been reported by collaborations such as CDMS~\cite{bib:cdms_si} and CoGeNT~\cite{bib:cogent}, as well as from gamma ray surveys in the galactic center~\cite{bib:hooper}. These excesses can be interpreted in terms of low mass WIMPs (5 - 30 GeV/c$^2$). These findings have renewed interests for light dark matter. Low mass WIMPs are particularly hard to identify in direct detection experiments because they generate extremely low energy recoils. This means a large part of the signal can be hidden by threshold effects. Event discrimination is also more difficult near the threshold because the different populations tend to overlap.    

\subsection{Data selection}

In the analysis shown here, we used only a small fraction of the whole data set shown in Fig.~\ref{fig:FID}. We unblinded a single, standard detector (called FID837) to tune the analysis and build background models. The analysis threshold was fixed at 3.6 keVnr (nuclear recoil energy scale). This is a conservative value, chosen such that the online trigger efficiency is 100\% and that threshold effects can safely be neglected. 
~\newline The data shown in this paper was first passed through general selection cuts. These include a period selection based on the online trigger level: periods where the online trigger is higher than 2.4 keVnr were rejected. We also discarded time periods with noisy ionization or heat signals in order to ensure that the data set remains homogeneous. We also include quality cuts to remove noisy or poorly reconstructed events. Since we do not expect WIMP multiple scatters, multiple events are rejected. We further discard pile-up events which can lead to erroneous interpretations. Taking into account the efficiency of all these cuts, the fiducial volume and the deadtime correction leads to a final, effective exposure of 35 kg.d.  
~\newline Then, we defined a loose "WIMP-box" cut in order to shrink the parameter space to the region with the highest likelihood of finding WIMPs. That is why we set the analysis range to 3.6~keVnr - 30 keVnr and imposed a lax cut on the veto electrodes (require less than 5 standard deviation on the veto electrodes). This cut has the advantage of facilitating the work of multivariate methods (which need to scan less parameter space) while leading to negligible efficiency loss.

\subsection{Background and Signal models}

~\\All background models in this analysis were data driven. We have used sideband (i.e. region without signal) data to feed a generative model for well-known backgrounds (gamma, beta, lead recoils). Surface event models were cross checked against calibration data and were found to be in good agreement with a Kolmogorov Smirnov test. The so-called heat-only background required a more careful study. As can be seen from Fig.~\ref{fig:results}, this is the dominant background for this analysis. It is made of events for which only the heat signal is clearly identified. These events chart an irregular rate over time and are likely due to mechanical vibrations. Fortunately, above the analysis threshold, heat-only events can be fully characterised in a sideband.

~\\The WIMP signal was generated using the well known theoretical formula for nuclear elastic scattering~\cite{bib:lewin&smith}. The parameterisation for the nuclear quenching factor is that of the EDELWEISS experiment, verified by many neutron calibrations: ${\rm Q(E)}=0.16\times {\rm E}^{0.18}$.

\subsection{Results}

\begin{figure}[!ht]
\centering
\includegraphics[height=6cm,width=0.49\textwidth]{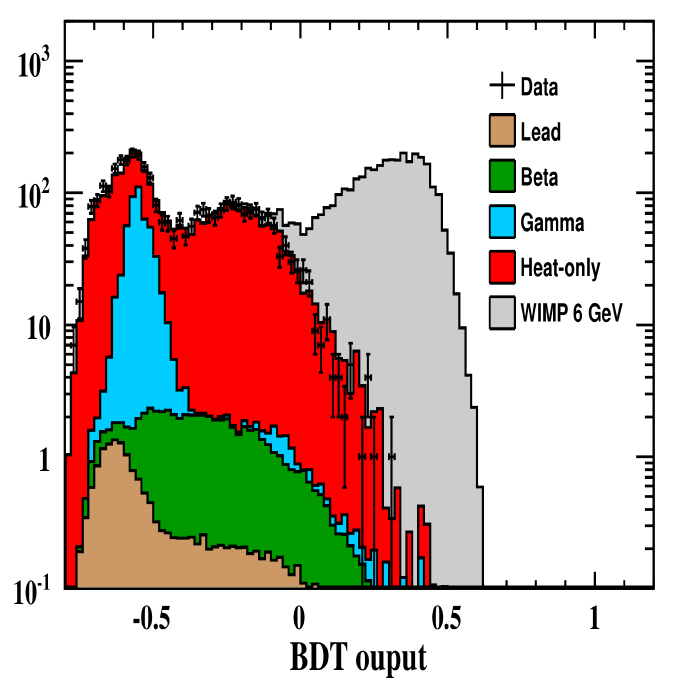}
\includegraphics[height=6cm,width=0.49\textwidth]{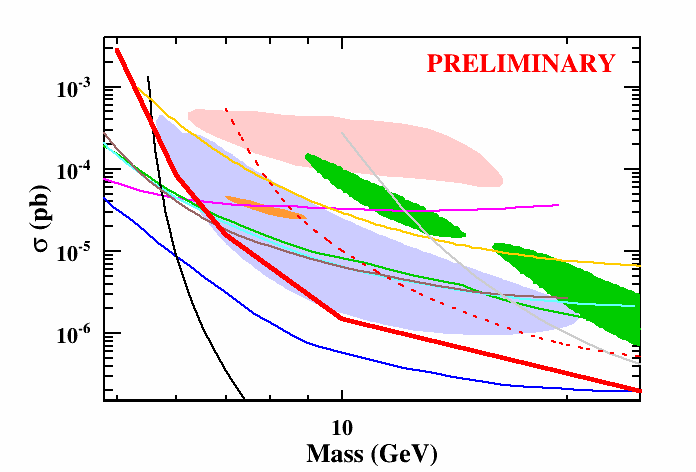}
\caption{{\bf Left:} Boosted Decision Trees' discriminating variable. The colored histograms show the background contributions, the grey histogram shows the expected WIMP signal from a 6~GeV WIMP and the black dots are the data. {\bf Right:} Limit on the WIMP cross section, given the WIMP masses. Color code: SCDMS ({\em blue}), CDMS-Si contour ({\em light blue}), CDMSlite ({\em purple}), DAMA ({\em salmon}), CRESST limit and contours ({\em green}), SIMPLE ({\em yellow}), COUPP ({\em gray}), PICO ({\em teal}), Xenon10 ({\em brown}), LUX ({\em black}), CoGeNT ({\em orange}), EDELWEISS-II ({\em dashed red}) and preliminary EDELWEISS-III 35 kg.d in ({\em red}) (this work).}
\label{fig:results}
\end{figure}

~\\We used TMVA's~\cite{TMVA2007} Boosted Decision Trees (BDT) for the event discrimination. This is a multivariate method which combines several inputs into a single discriminating variable. The BDT first undergoes a training phase where it learns to separate background from signal. The feature space is spanned by the 4 ionisation and 2 heat channels of the EDELWEISS experiment. The training sample is generated with the background and signal models outlined in the previous section. Given that this work is essentially a validation study for future analyses, we decided to report an upper limit on the WIMP cross section. This was done by defining an optimal cut value on the BDT output and quoting the upper limit corresponding to the number of observed events after cut. The optimal cut value was derived from further simulations by maximising the signal over noise ratio, effectively rejecting all backgrounds ($<1$ background event expected). A BDT was trained for each WIMP mass. The resulting limit is shown in Fig.~\ref{fig:results} (right), showing competitive results in spite of the small exposure and relatively high threshold. Fig.~\ref{fig:results} (left) shows that a clear separation between signal and background events can be achieved. This is a tribute to the new FID detector design which allows for remarkable surface event rejection. This clearly demonstrates the potential of EDELWEISS bolometers for low mass WIMP searches.

\section{Conclusions}

We analyzed the first data from the EDELWEISS-III experiment in a low mass WIMP search. The results are very promising for future searches: improvements in the baseline resolution (and hence the experimental threshold) allow a single detector (35 kg.d of exposure) to beat the published EDELWEISS-II low mass limit~\cite{bib:edw2} (113 kg.d.)
The experimental sensitivity will further increase by pushing the analysis in two directions: we are going to increase the available statistics by combining several detectors and we plan to decrease the analysis threshold in order to improve the sensitivity to very low mass WIMPs ($<5$~GeV/c$^2$).

\newpage

\section{Acknowledgements}

The help of the technical staff of the Laboratoire Souterrain de Modane and the participant laboratories
is gratefully acknowledged. The EDELWEISS project is supported in part by the German
ministry of science and education (BMBF Verbundforschung ATP Proj.-Nr. 05A14VKA),
by the Helmholtz Alliance for Astroparticle Phyics (HAP),
by the French Agence Nationale pour la Recherche and the Labex Lyon Institute of Origins (ANR-10-LABX-0066)
of the Université de Lyon within the program "Investissement d'Avenir" (ANR-11-IDEX-00007),
by Science and Technology Facilities Council (UK) and the Russian Foundation for Basic Research (grant No. 07-02-00355-a).

\end{document}